\documentclass[preprint,12pt]{elsarticle}

\journal{Physica A}

\usepackage{lineno,hyperref}
\usepackage{graphics}
\usepackage{anyfontsize}

\usepackage{graphicx}	
\usepackage{amssymb}
\usepackage{color}
\usepackage{xcolor}  
\biboptions{sort&compress}
\usepackage{hyperref}  
\usepackage{multirow}
\hypersetup{colorlinks,urlcolor=blue,linkcolor=blue,citecolor=blue,filecolor=blue,urlcolor=blue}

\begin{document}

\begin{frontmatter}



\title{Portfolio Theory, Information Theory and Tsallis Statistics}

\author[mymainaddress]{Marco A. S. Trindade}

\author[mysecondaryaddress]{Sergio Floquet\corref{mycorrespondingauthor}}
\cortext[mycorrespondingauthor]{\\ \phantom{aa}* Corresponding author}
\ead{sergio.floquet@univasf.edu.br}

\author[mymainaddress]{Lourival M. Silva Filho}

\address[mymainaddress]{Colegiado de F\'isica, Departamento de Ci\^encias Exatas e da Terra, Universidade do Estado da Bahia, Brasil}
\address[mysecondaryaddress]{Colegiado de Engenharia Civil, Universidade Federal do Vale do S\~ao Francisco, Brasil}

\linenumbers

\begin{abstract}
We developed a strategic of optimal portfolio based on information theory and Tsallis statistics. The growth rate of a stock market is defined by using $q$-deformed functions and we find that the wealth after n days with the optimal portfolio is given by a $q$-exponential function. In this context, the asymptotic optimality is investigated on causal portfolios, showing advantages of the optimal portfolio over an arbitrary choice of causal portfolios. Finally, we apply the formulation in a small number of stocks in brazilian stock market  $[B]^{3}$ and analyzed the results. 
\end{abstract}


\end{frontmatter}


\section{Introduction}

   The Modern Portfolio Theory 
   was introduced by Harry Markovitz in 1952 \cite{Mark}. The basic principle is the mean-variance approach \cite{Sharpe} so that the expected return is maximized for a given level of risk, i.e., a constraint on the variance. This reflects the idea of diversification in investment and risk aversion. Posteriorly, Cover explored these concepts in the context of information theory \cite{Cover,Cover1,Cover2,Cover3}. Particularly, Kelly introduced the concept of log-optimal portfolio \cite{Kelly}. An asymptotic equipartition property for the stock market as well as the asymptotic optimality of log-optimal investment was derived by Algoet and Cover \cite{Cover4}. Still in this context of quantitative finance, the Cover's portfolio 
   was defined in reference \cite{Cover5}. 
   
 It is important to highlight two seminal references in the applications of information theory in finance \cite{theil65,fama65b}. In \cite{theil65} Theil and Leenders analyzed data on the Amsterdam Stock Exchange in the period November 2, 1959, through October 31, 1963, concluding that the Amsterdam Stock Exchange has a memory of one day declining in price, advancing or remaining unchanged. The relative entropy was used as a measure of inaccuracy  for the forecasts. Fama \cite{fama65b}, in turn, applied the Theil-Leenders tests in data from the New York Exchange for the period June 2, 1952, to October 29, 1962, conducting that the proportions of securities declining or advancing today on the New York Stock Exchange are not auspicious in predicting proportions declining or advancing  tomorrow.

 A relevant risk measure in portfolio optimization is the entropic value-at-risk (EVarR), introduced by Ahmadi-Javid \cite{ahmadi11,ahmadi12}. In addition to being coherent (i.e., it satisfies the properties translation invariance, subadditivity, monotonicity, positive homogeneity), it is strongly monotone and strictly monotone. Recently, a promising sample-based portfolio optimization has been proposed \cite{ahmadi19}. An interesting approach is that EVaR approach outperforms CVaR (Conditional Value-at-Risk) approach as the sample size increases. This is because the number of variables and constraints of the EVaR is independent of the sample size. In addition, under a analysis of real have better best mean and worst return rates and Sharpe ratios compared  with the previous one.  An investigation about  performance hypothesis testing through the Sharpe ratio was conducted by Memmel \cite{memmel03}. It has been shown that we can use as test statistic the Sharpe ratio difference divided by its asymptotic standard deviation. 
Aifan Ling et al. \cite{ling19} developed a robust multi-period mean-LPM (lower partial moment) portfolio selection model considering transaction cost under an asymmetric uncertainty set. This model provides better returns an Shape ratios when real market data are analyzed. 
%

A recent approach for the field of quantitative finance and economics, called econophysics \cite{St1}, lies in the use tools derived from statistical physics \cite{Bou}. Relations between physics and economics are long standing. The gravity model of international trade, for example, mimics the law of gravitation and it was proposed in 1954, by Walter Isard \cite{Isard}. In the econophysics framework associated to information theory, some works \cite{Zu1, Zu2} were developed exploring the complexity-entropy causality plane. In the reference \cite{Zu1}, applications to distinguish the stage of the development of the stock market are conducted and in \cite{Zu2} the efficiency of sovereign bond markets is investigated through complexity-entropy causality plane revealing correlations and hidden structures in the daily values of bond indices.
Zunino et al. \cite{zunino16} analyzed the time-varying informational efficiency of European corporate bond markets as well as the impact of the 2008 financial crises on setorial indices related to the aforementioned titles. An interesting result is that before the crisis, all sectors present similar efficient behaviors. In the post-crisis each sector follow  its  own dynamic.

Another techniques emerge from non-extensive statistical mechanics \cite{Tsallis2, Tsallis3} proposed by Tsallis in 1988 \cite{Tsallis1}, as a generalization of Boltzmann-Gibbs statistics. The Tsallis entropy carries a non-extensive parameter q such that in the limit $q \rightarrow 1$ the Boltzmann-Gibbs entropy is recovered. The maximization of the Tsallis entropy under appropriate constraints leads to the distribution of Tsallis $q$-Gaussian \cite{Tsallis2, Tsallis3, Tsallis4}. It has been widely used in the study of complex systems \cite{Tsallis2, Tsallis3}, including the stock market.%

The $q$-Gaussians allow a consistent way of describing high-frequency financial observations due to finite variance and temporal autocorrelations \cite{TsBor}. Option pricing formulas based on Tsallis statistics were derived by Borland \cite{Borland}. As highlighted in \cite{TsBor}, one advantage is the existence of explicit closed-form solutions. The generalized Black-Sholes (B-S) equation was obtained with entropic index $q$. This model fits, for $q=1.5$, stock returns more realistically than B-S standard $(q=1)$. \

The probability distribution of stock returns is non-Gaussian \cite{mandelbrot63,fama65}. On the other hand, the modern portfolio theory assumes that returns follow a Gaussian distribution, which results in a less realistic scenario. In order to overcome these limitations,  in this paper, we merge Cover's approach from information theory to 
the mathematical tools of Tsallis statistics. In section \ref{sec1} we present the Nonextensive Statistics and some of yours results that will be using in section \ref{sec2} to formulate the nonextensive version of the Cover's portfolio   using the $q$-deformed functions and the $q$-product as key elements. In addition, an asymptotic study is carried out to substantiate the work. In section \ref{sec3} we will apply the above for-mentioned formalism for the analysis of a small number of stocks in brazilian stock market and comparing the results with Cover's portfolio describe by a Gaussian distribution, besides calculating the Sharpe Ratio and Sortino Ratio. Finally, we present the final considerations and perspectives in section \ref{sec4}.

\section{Nonextensive Statistics \label{sec1}} 

\hspace{0.5cm} Tsallis entropy \cite{Tsallis1} is a generalization of the Boltzmann-Gibbs entropy. It is given by
\begin{eqnarray}
 S_{q} & = & k_{B} \frac{1- \displaystyle \sum_{i=1}^{n} p_{i}^q }{1-q} , 
\end{eqnarray}
with  $\displaystyle \sum_{i=1}^{n} p_{i} = 1$, where for two systems independent $A$ and $B$ we have the propriety
\begin{eqnarray}
 S_{q}(A+B) = S_{q}(A) + S_{q}(B) +(1-q)S_{q}(A)S_{q}(B).
\end{eqnarray}
The parameter $q$ represents the nonextensive of the systems, when for $q=1$ we recover the Boltzmann-Gibbs entropy, for $q<1$ we obtain the superextensivity (additivity of the entropy) and for $q>1$ we obtain the subextensivity (decrease of the entropy). 

Tsallis introduced the function $q$-logarithm  and $q$-exponential
\begin{eqnarray}
\ln_{q}(x) &=& \displaystyle \frac{ x^{1-q}-1}{1-q}, \ \ \exp_{q}(x)  =  [1+(1-q)x]_{+}^{\frac{1}{1-q}} \label{eq-lnq}
\end{eqnarray}
with $[A]_{+}:=max(0,A)$, that for $q=1$ we recover the logarithm and exponential functions, and allow write Tsallis entropy as 
$S_{q} = -\sum_{i=1}^{n} p_{i}^{q} \ln_{q}p_{i}$. Another important operation between the functions is the $q$-product, $\otimes_{q}$, that is defined by \cite{Borges}
\begin{equation}
x \otimes_{q} y := [x^{1-q}+y^{1-q}-1]_{+}^{\frac{1}{1-q}}. \label{eq-q-produto}
\end{equation}

The $q$-Gaussian function can be introduced as
\begin{eqnarray}
f_{q}(x) & = & \frac{1}{C_{q}|\sigma|}  exp_{q}\left[ -\frac{(x-\mu)^{2} }{\sigma^{2}} \right]
\end{eqnarray}
where $C_{q}$ is the normalization constant given by
\begin{equation}
 C_q=\left\{	\begin{array}{ll}
  \frac{2\sqrt{\pi}\Gamma \left(\frac{1}{1-q}\right)}{(3-q)\sqrt{1-q}\Gamma \left(\frac{3-q}{2(1-q)}\right)}&
   -\infty<q<1 \\
		\vspace{0.1cm}
    \sqrt{\pi},&  \ \ q=1 \\
		\vspace{0.1cm}
    \frac{\sqrt{\pi}\Gamma \left(\frac{3-q}{2(q-1)}\right)}{\sqrt{q-1}\Gamma \left(\frac{1}{1-q}\right)} ,&   1<q<3, \\
  \end{array}	\right.
  \end{equation}
which has an asymptotic ($x\gg 1$) power law behavior given by $ \exp_{q}(x^{2}) \sim  x^{ \frac{2}{1-q}}$. In a similar way we have the multivariate $q$-Gaussian distribution
\begin{eqnarray} \label{qGau}
f_{q}(\textbf{x})&=&C_{d,q}\exp_{q}\left[ \frac{ (x_{1}-\mu_{1}) ^{2} }{\sigma_{1}^{2}} + \cdots + \frac{ ( x_{d} -\mu_{d} ^{2}) }{\sigma_{d}^{2}} \right],
\end{eqnarray}
where $\sigma_{i}$ and $\mu_{i}$ are parameters to be determined and the normalizing constant is given by
 $ C_{d,q}  =  \displaystyle \frac{   1  }{ |\sigma_{1}\cdots \sigma_{d}| w_{d} I_{q,d} }$ where $w_{d}$ is the surface area of the unit sphere in $\mathbb{R}^{d}$ dimensional space, with $ \displaystyle w_{d} = \frac{  2\pi^{ \frac{d}{2} }   }{ \Gamma(\frac{d}{2}) } $, and $ \displaystyle I_{q,d}  =  \int_{0}^{+\infty} r^{d-1} e_{q}^{-r^{2}} $.
 
Umarov and Tsallis  \cite{Umarov1} obtained a formula for compute the normalizing constant
\begin{eqnarray}
C_{d,q}= \frac{\left(\frac{3-q_{1}}{2}\right)^{\frac{d-1}{2}}\left(\frac{3-q_{2}}{2}\right)^{\frac{d-2}{2}}...
\left(\frac{3-q_{d-1}}{2}\right)^{\frac{1}{2}}}{(C_{q}^{q})^{-1}(C_{q_{1}}^{q_{1}})^{-1}...(C_{q_{d-1}}^{q_{d-1}})^{-1}},
\end{eqnarray}
with
\begin{eqnarray}
q_{n}=\frac{2q+n(1-q)}{2+n(1-q)}, \ \ n=0 \pm 1, \pm 2,...
\end{eqnarray}
where $C_{q}$ is the normalizing constant of one-dimensional $q$-Gaussian.

As pointed out in many papers \cite{TsBor,borland2004a,queiros2005a,vellekoop2007} the stock market data express a fat tail behavior and a power law for the cumulative return distribution in the asymptotic case ($x\gg 1$) \cite{drozdz2007b,pan2008,eryigit2009} what makes the $q$-Gaussian distribution attractive to describe stock market data. In cases \cite{borland2004a,gopikrishnana1998,drozdz2007a,ruiz2018} an inverse cubic power law for cumulative distribution was obtained and it is related to $q$-Gaussian distribution with $q=1.5$.

\section{Cover's q-Portfolio \label{sec2}} 
The stock market is given by a vector $\mathbf{X}=(X_{1}, X_{2},..., X_{n})$, where $X_{i}$ is the relative price, i.e. the ratio of the price at the end of the day to the price at the beginning of the day. We have that $X_{i}\geq 0$, $i=1,...,m$ where $m$ is the number of stocks. In this work  we apply the Tsallis statistic formalism in the Cover's portfolio theory, where the parameter $q$ is responsible by take account the nonextensivity of stock market. Following Cover \cite{Cover5} we introduced   the growth $q$-rate of a stock market portfolio $\textbf{b}$:
\begin{eqnarray}
W_{q}(\textbf{b}, f)&=&E(\ln_{q}\textbf{b}^{t}\textbf{x}) \nonumber \\
                    &=&\int \ln_{q}(\textbf{b}^{t}\textbf{x})f_{q} (\textbf{x})d\textbf{x}.
\end{eqnarray}
 We define the optimal growth $q$-rate $W_{q}^{*}(f_{q})$ as
\begin{eqnarray}
W^{*}_{q}(f)=\max_{\textbf{b}} W_{q}(\textbf{b}, f_{q}).
\end{eqnarray}
The growth optimal portfolio $b^{*}$ is one that achieves the maximum of $W_{q}(\textbf{b}, f_{q})$. Let $\textbf{X}_{1},\textbf{X}_{2},...,\textbf{X}_{n}$ be random vectors \emph{i.i.d} with density probability function $f_{q}$. The $q$-wealth after n days using the portfolio $b^{*}$ is given by
\begin{eqnarray}
S_{n}^{*(q)}=\bigotimes_{(q);i=1}^{n} \textbf{b}^{*t} \textbf{X}_{i},
\end{eqnarray}
where the $q$-product $\otimes_{q}$ is defined in (\ref{eq-q-produto}).

Using the strong law of large numbers \cite{Durrett}, we have
\begin{eqnarray}
\frac{1}{n}\ln_{q}S_{n}^{*(q)}&=&\frac{1}{n}\ln_{q}\bigotimes_{(q);i=1}^{n}\textbf{b}^{*t} \textbf{X}_{i} \nonumber \\
                              &=& \frac{1}{n}\sum_{i=1}^{n}\ln_{q}\textbf{b}^{*t} \textbf{X}_{i} \rightarrow W^{*}_{q}  \ \ a.s.,
\end{eqnarray}
so that $S_{n}^{*(q)}=\exp_{q}(nW^{*}_{q})$. This justifies our definition of $q$-wealth analogously to reference \cite{Cover}.
Now, we will consider the following assumption:
\begin{equation}\label{des1}
E \left(\frac{S_{n}^{(q)}}{S_{n}^{*(q)}}\right)\leq 1,
\end{equation}
in order to show the asymptotic optimality of the $ln_{q}$-optimal portfolio for the causal portfolios. The causal portfolio strategy is a sequence of mappings $b_{i}: T^{m(i-1)}\rightarrow B$, where the portfolio $\textbf{b}_{i}(\textbf{X}_{1},\textbf{X}_{2},..., \textbf{X}_{i-1})$ is used on day $i$. We define $B=\{\textbf{b} \in \mathbb{R}^{m}: \textbf{b}_{i} \geq 0, \sum_{i=1}^{m}\textbf{b}_{i}=1 \}$ as the set of allowed portfolios. Using the Markov inequality \cite{Durrett}, we have from (\ref{des1}):
\begin{eqnarray}
P(S_{n}^{(q)} > \lambda_{n}S_{n}^{*(q)}) \leq \lambda_{n}^{-1}.
\end{eqnarray}
Therefore
\begin{eqnarray}
P\left(\frac{1}{n}\ln_{q}\frac{S_{n}^{(q)}}{S_{n}^{*(q)}} > \frac{\ln_{q}\lambda_{n}}{n} \right) \leq \lambda_{n}^{-1}.
\end{eqnarray}
Consequently
\begin{eqnarray}
\sum_{n=1}^{\infty} P\left(\frac{1}{n}\ln_{q}\frac{S_{n}^{(q)}}{S_{n}^{*(q)}} > \frac{n^{2-2q}-1}{n(1-q)} \right) \leq \frac{\pi^{2}}{6},
\end{eqnarray}
where we consider $\lambda_{n}=n^{2}$. Then
\begin{eqnarray}
P\left(\frac{1}{n}\ln_{q}\frac{S_{n}^{(q)}}{S_{n}^{*(q)}} > \frac{n^{2-2q}-1}{n(1-q)} ,\textrm{infinitely often} \right)=0
\end{eqnarray}
using the Borel-Cantelli lemma \cite{Durrett}. Thus, there exists an $N$ such that for all $n>N$:
\begin{equation} \label{alt}
\frac{1}{n}\ln_{q}\frac{S_{n}^{(q)}}{S_{n}^{*(q)}} \leq \frac{n^{2-2q}-1}{n(1-q)},
\end{equation}
for almost every sequence from the stock market. This implies (for $q>0.5$) with probability 1:
\begin{equation}
\limsup_{n \rightarrow \infty} \frac{1}{n} \ln_{q}\frac{S_{n}^{(q)}}{S_{n}^{*(q)}} \leq 0.
\end{equation}

We have that for almost every sequence from stock market, $S_{n}^{*(q)}$ is greater than the wealth of any investor, i. e., the $ln_{q}$-optimal portfolio is better than any other portfolio under the assumptions aforementioned.  We can also show that
\begin{eqnarray}
\max_{\textbf{b}_{1},...,\textbf{b}_{n}}E \left( \ln_{q}S_{n}^{(q)} \right) &=&\max_{\textbf{b}_{1},...,\textbf{b}_{n}}E \left[ \ln_{q}\left( \bigotimes_{(q);i=1}^{n} \textbf{b}_{i}^{t}\textbf{X}_{i} \right) \right] \nonumber \\
&=&\sum_{i=1}^{n}\max_{\textbf{b}_{1},...,\textbf{b}_{n}}E \left( \ln_{q}\textbf{b}_{i}^{t}
(\textbf{X}_{1},...,\textbf{X}_{i}) \right) \nonumber \\
&=&\sum_{i=1}^{n}E \left( \ln_{q}\textbf{b}^{*t}\textbf{X}_{i} \right) \nonumber \\ 
&=& nW^{*}_{q}.
\end{eqnarray}
Consequently,
\begin{eqnarray}
E \left( \ln_{q}S_{n}^{*(q)} \right) & \geq & E \left( \ln_{q}S_{n} \right) ,
\end{eqnarray}
i.e., the $\ln_{q}$-optimal portfolio maximizes the expected $\ln_{q}$ of the final wealth. So we show theoretically that the q-Portfolio will provide higher relative wealth results that the Cover's portfolio and, for $q \rightarrow 1$,  we recover the Cover's portfolio theory from the q-Portfolio.

\section{Computational results \label{sec3}}

In this section, we apply our formulation in the Brazilian stock market  $[B]^{3}$ - Brazil, Stock Exchange and Over-the-Counter Market located at S\~ao Paulo - Brazil, using \textit{R} language and environment \cite{r-cran}.
We use the package GetHFData \cite{gethfdata} that download and aggregate high frequency data from Brazilian stock market using File Transfer Protocol (FTP) and the package DEoptim \cite{deoptim} that implements the Differential Evolution Algorithm \cite{diffevol}. This is a useful method for solution of global optimization problems. Also, we use the packages Pracma, Cubature and R2Cuba  \cite{pracma,cubature,becs} that compute numerical multi-dimensional integration from Gauss-Kronrod, hypercubes and Monte Carlo methods, respectively.

Importing the Brazilian stock market data from the period 01/01/2018  to  04/30/2018, which represent 80 days of movement in $[B]^{3}$, we choose analyze the stocks with higher trading in $[B]^{3}$ as: Brazilian Petroleum Corporation - Petrobas (PETR4), Vale S.A. (VALE3), Bank of Brazil (BBAS3) and Bradesco Bank (BBDC4).

The Gaussian distribution was used in the Cover's portfolio theory  \cite{muller88,inuiguchi2000,li2013,tunc2013,matesanz08,markowitz12}, as well as $q$-Gaussian distribution \cite{zhao2018}. To apply our formalism describe in section \ref{sec2} we choose a multivariate $q$-Gaussian as joint probability density \ distribution \ of  the vector \ of price relatives, as  
defined in (\ref{qGau}), and   
compare this results with the multivariate Gaussian distribution for the Cover's portfolio theory related to relative prices \cite{Cover5}.

The parameters $q$, $\mu_{i}$ and $\sigma_{i}$ are estimated through maximum likelihood method for each set of stocks, and then the growth rate is maximized in order to construct the Cover's portfolio for the Gaussian distribution and $q$-Portfolio.

\begin{figure}[!htb]

\vspace{-2.8cm}\hspace{-2.5cm} \resizebox{1.4\textwidth}{!}{ \includegraphics{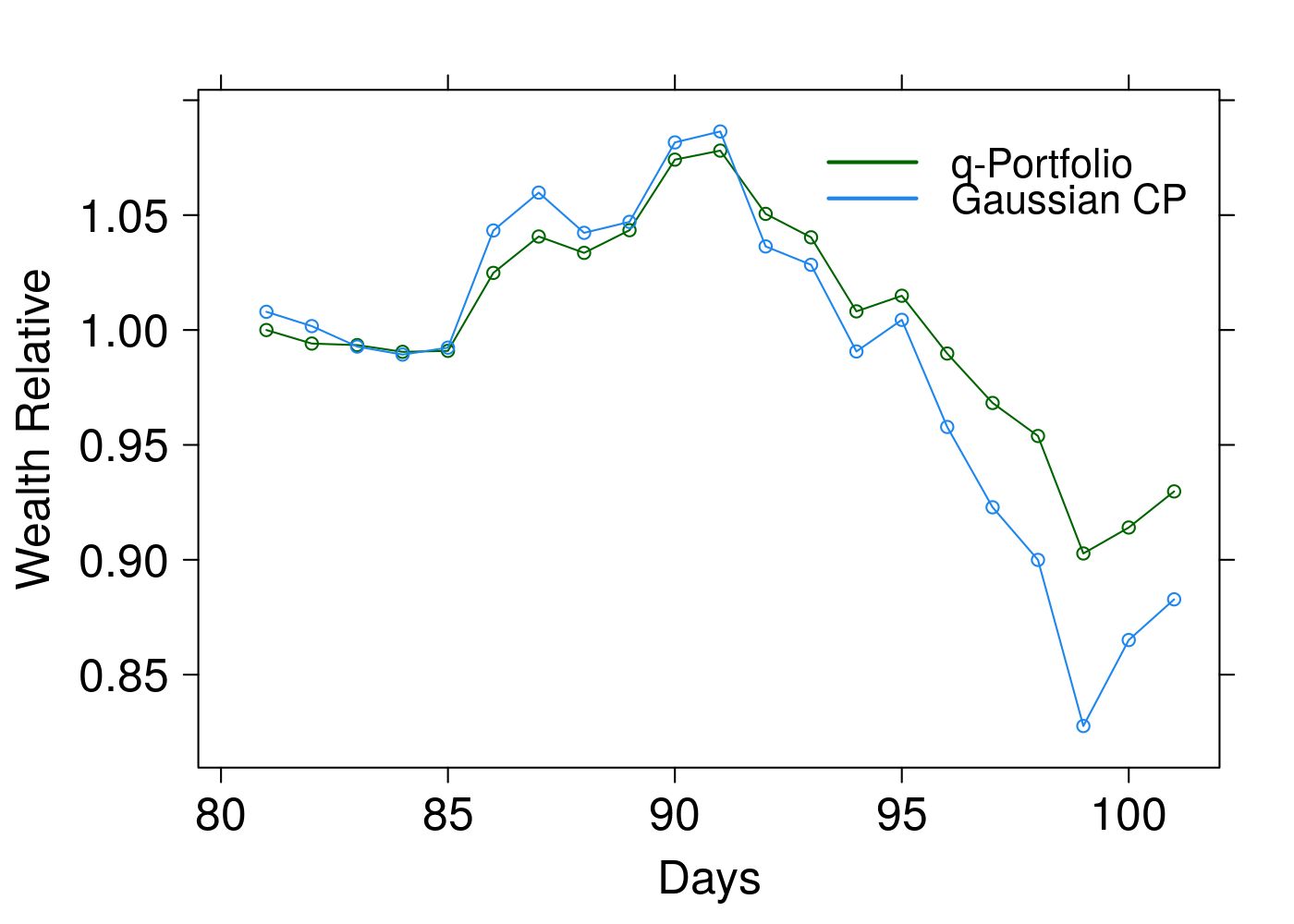}  
\includegraphics{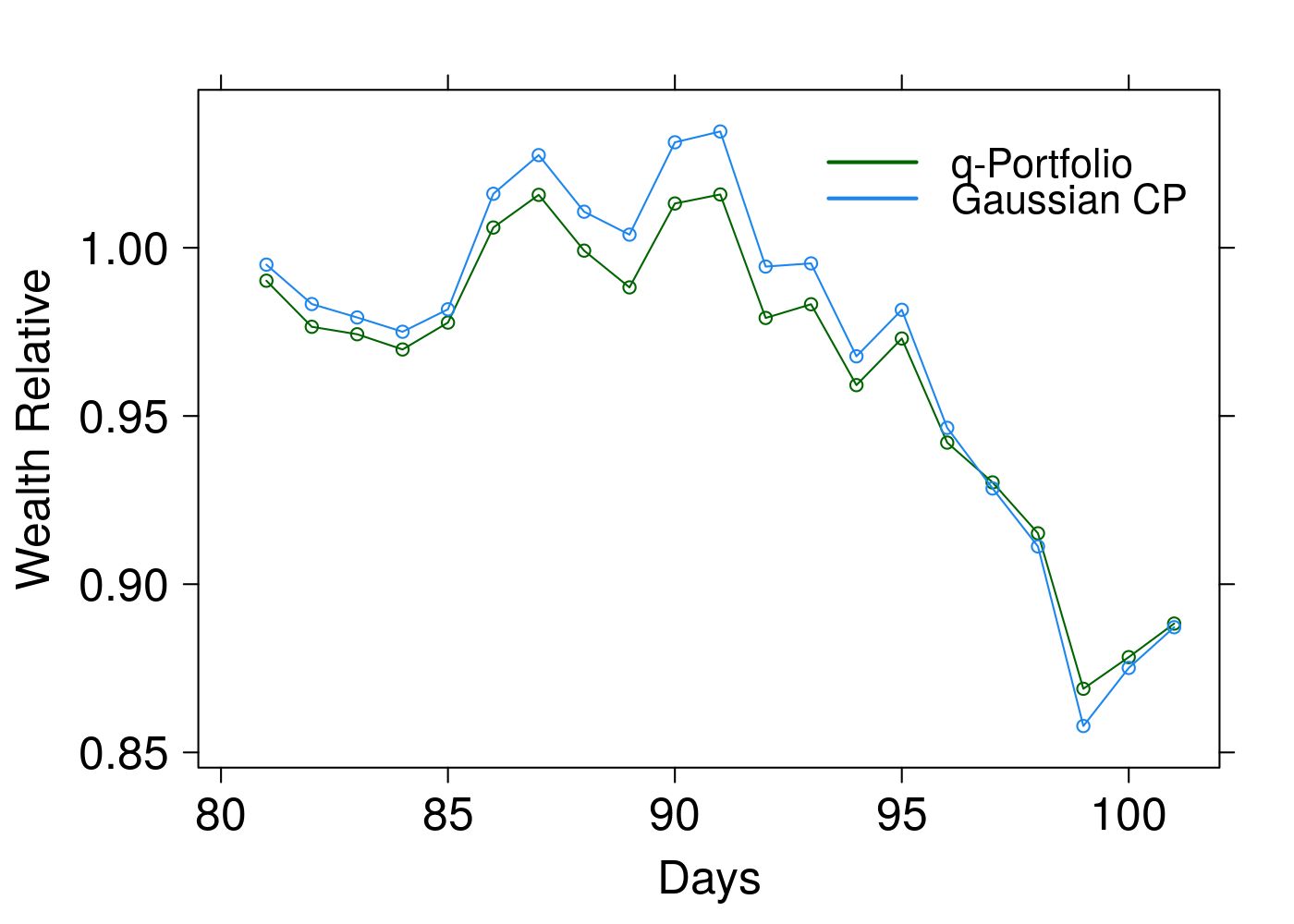} } 

\vspace{-0.7cm}\hspace{1.6cm}(a) \hspace{8.7cm} (b)

\hspace{-2.5cm} \resizebox{1.4\textwidth}{!}{ \includegraphics{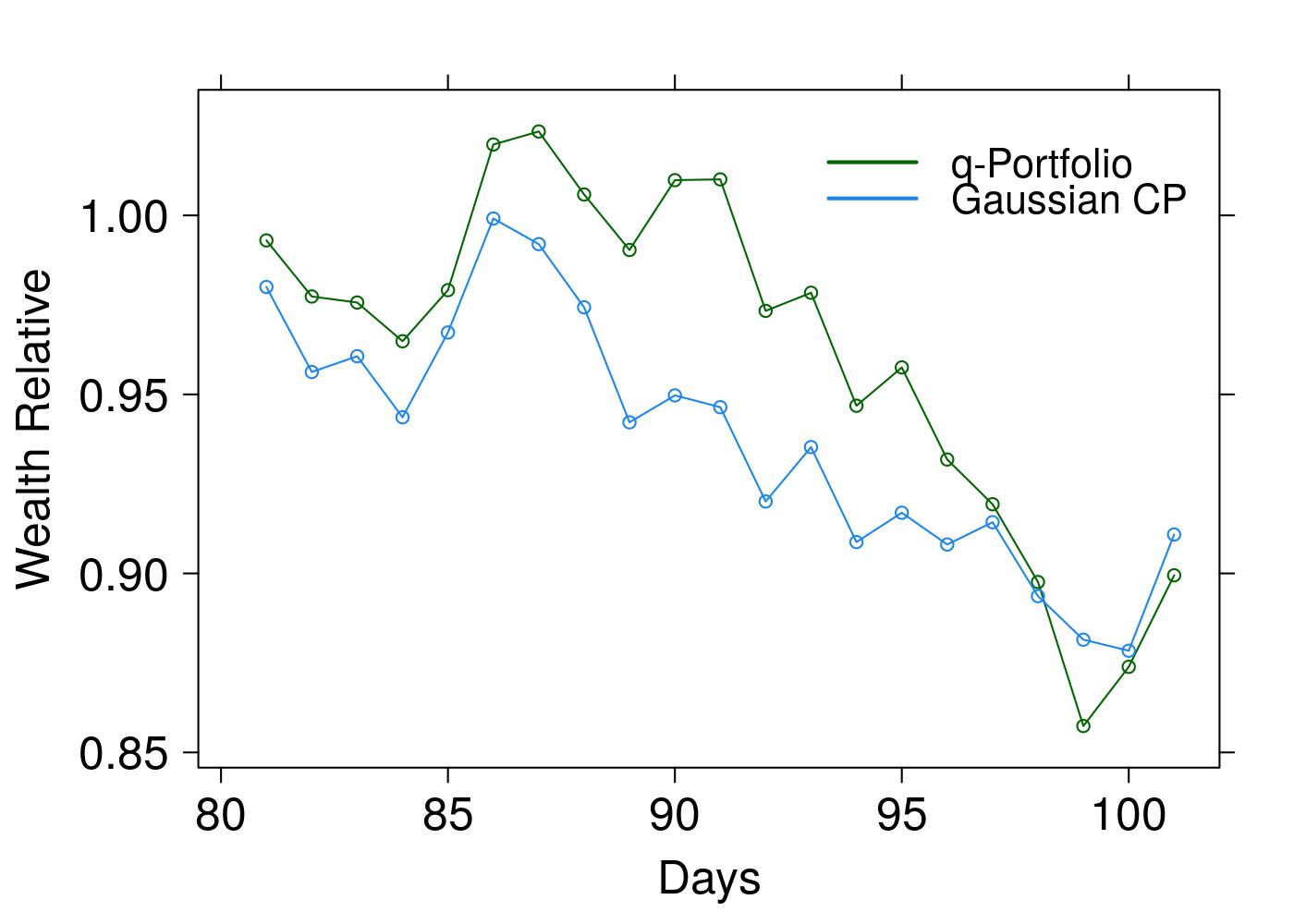}
\includegraphics{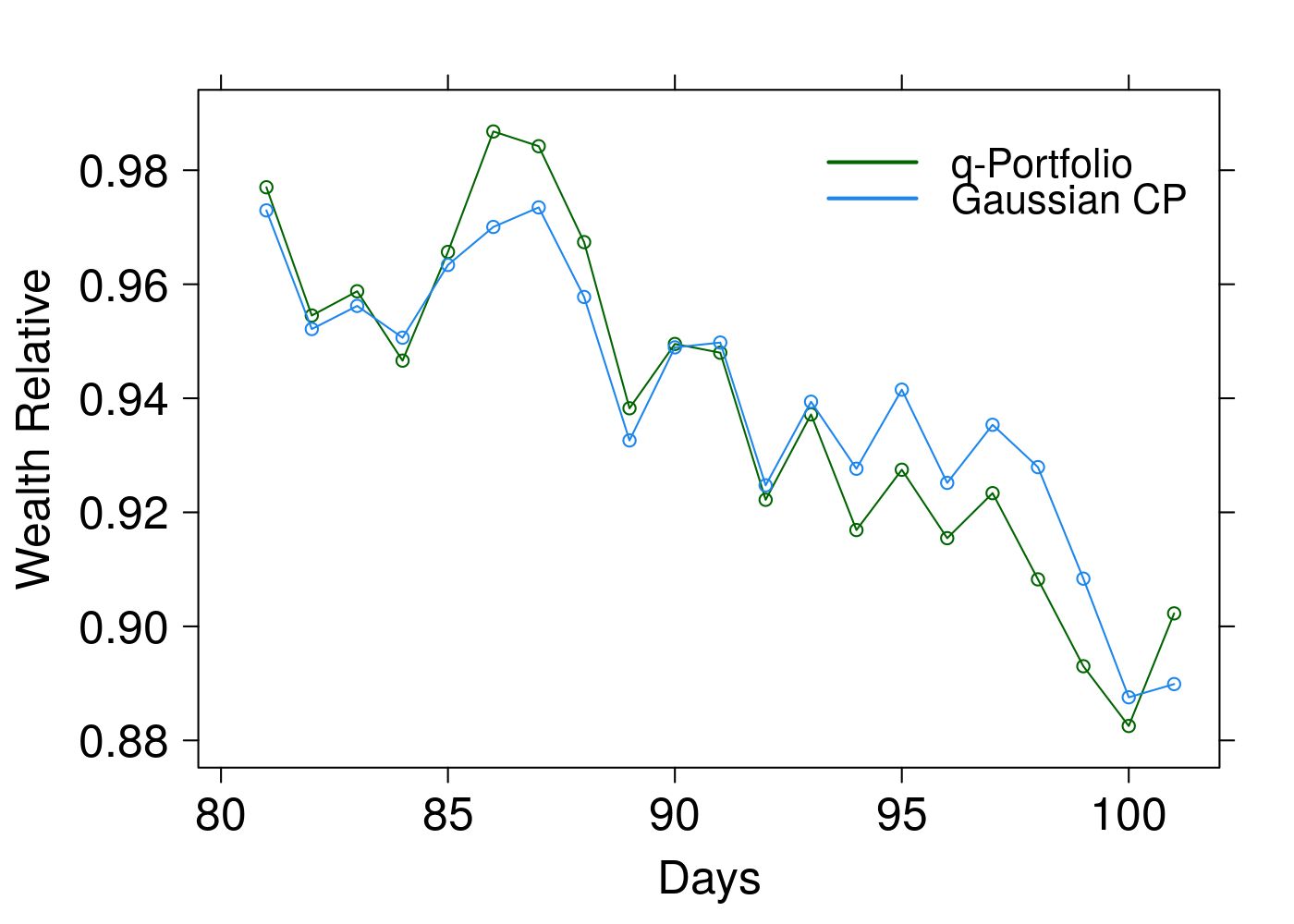} }

\vspace{-0.7cm}\hspace{1.5cm} (c) \hspace{8.7cm} (d)

\hspace{-2.5cm} \resizebox{1.4\textwidth}{!}{ \includegraphics{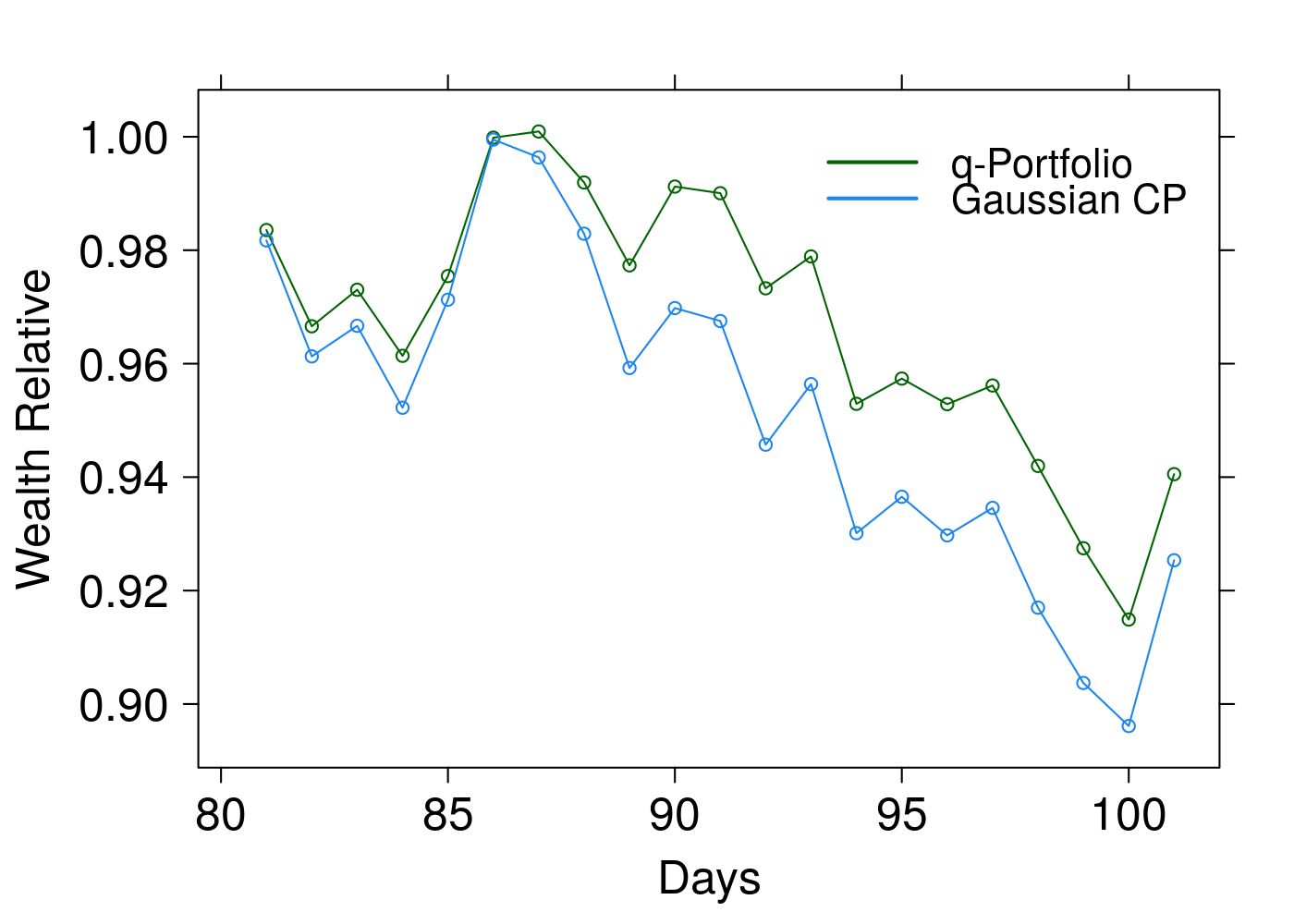}
\includegraphics{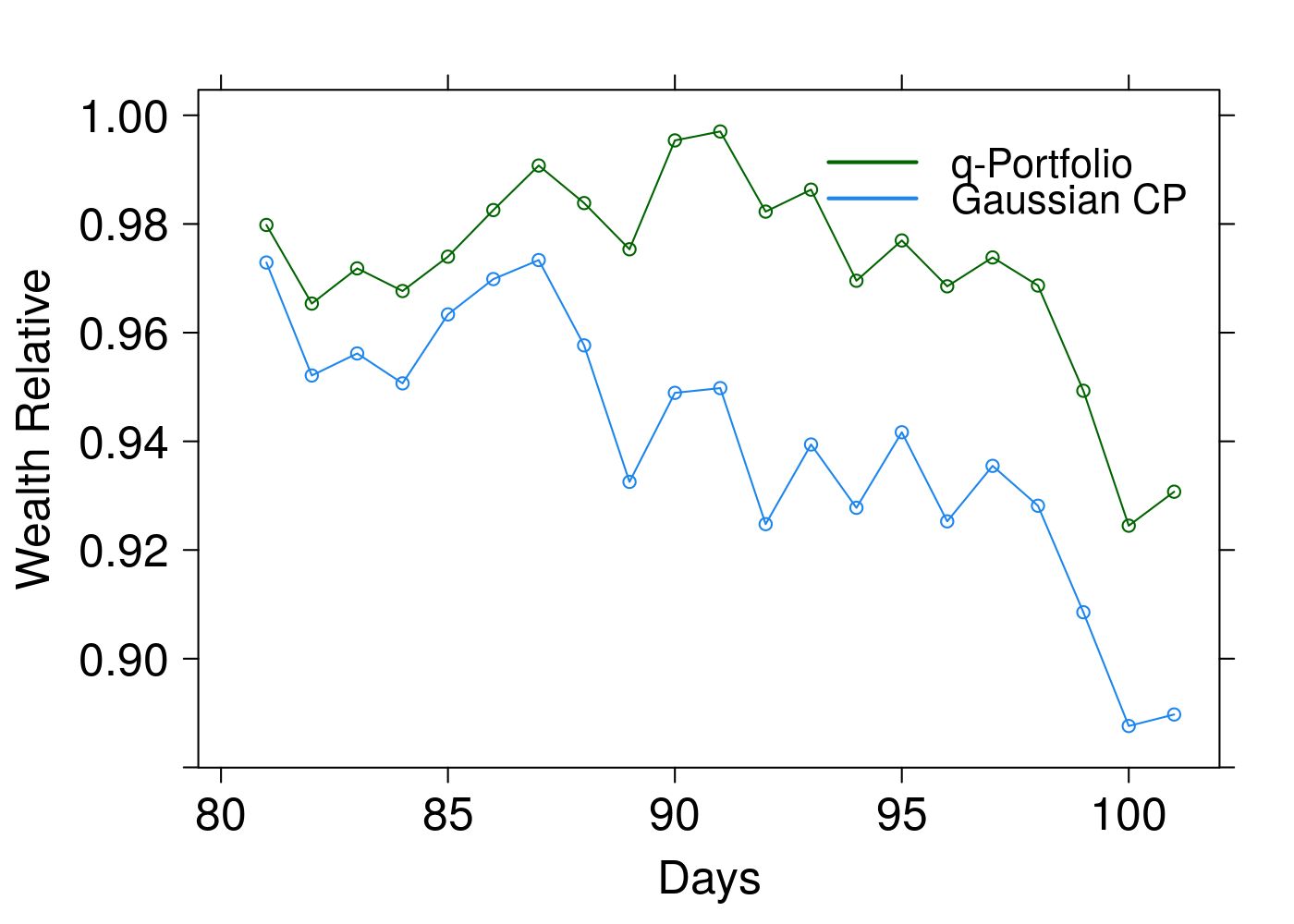} }

\vspace{-0.7cm}\hspace{1.5cm} (e) \hspace{8.7cm} (f) 

\caption{Wealth relative for the $q$-portfolio and the Gaussian Cover's Portfolio (CP) in the period 05/01/2018 to 05/31/2018, for two stocks: a) PETR4 and VALE3, b) BBDC4 and PETR4, c) BBAS3 and PETR4, d) BBAS3 and BBDC4, e) BBAS3 and VALE3, f) BBDC4 and VALE3.   }  \label{fig-2}    \vspace{-0.8cm}

\end{figure}

\begin{figure}[!htb]

\vspace{-2.8cm}\hspace{-2.5cm}  \resizebox{1.4\textwidth}{!}{ \includegraphics{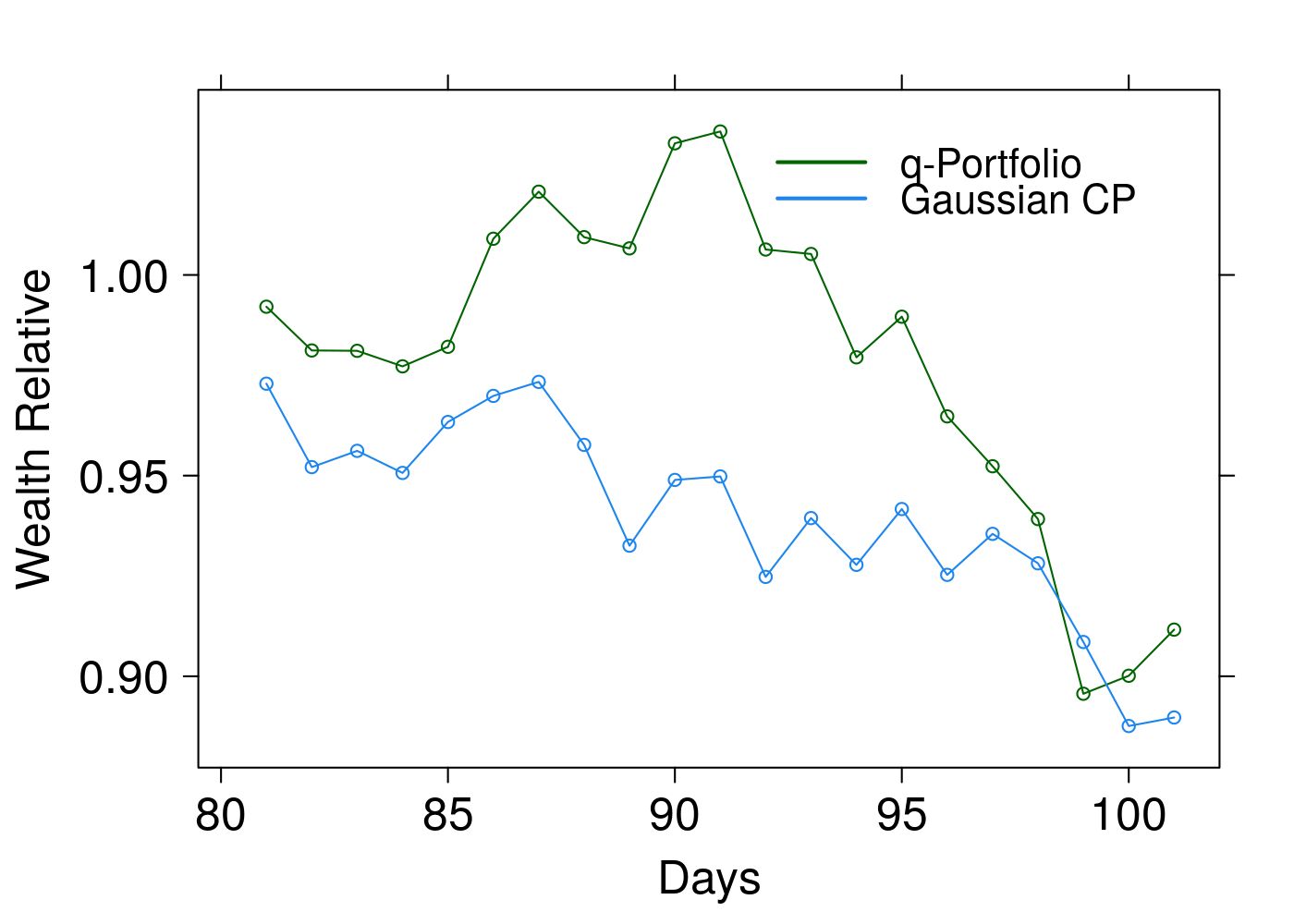}  
\includegraphics{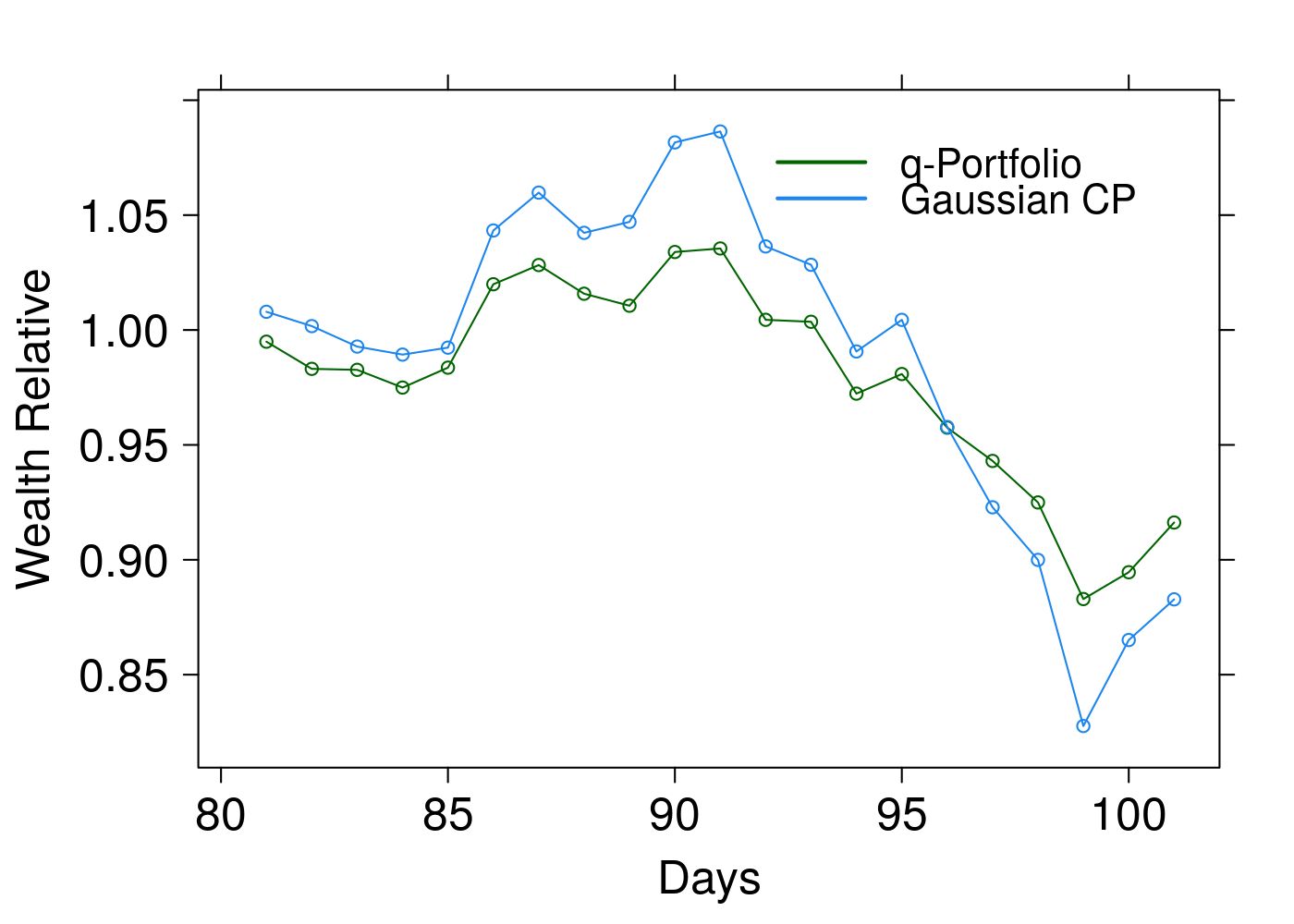} } 

\vspace{-0.7cm}\hspace{1.6cm}(a) \hspace{8.7cm} (b)

\hspace{-2.5cm}  \resizebox{1.4\textwidth}{!}{ \includegraphics{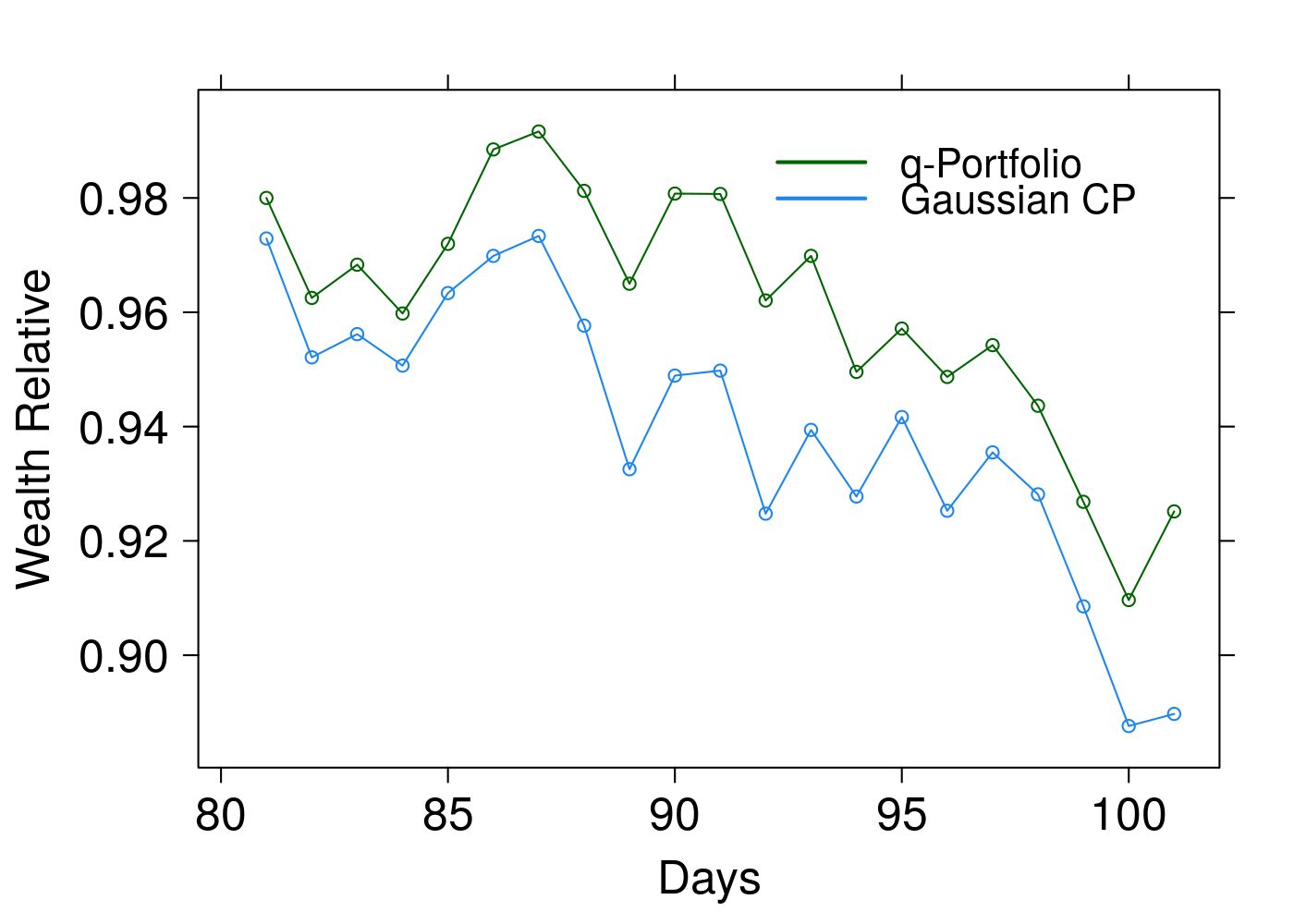}
\includegraphics{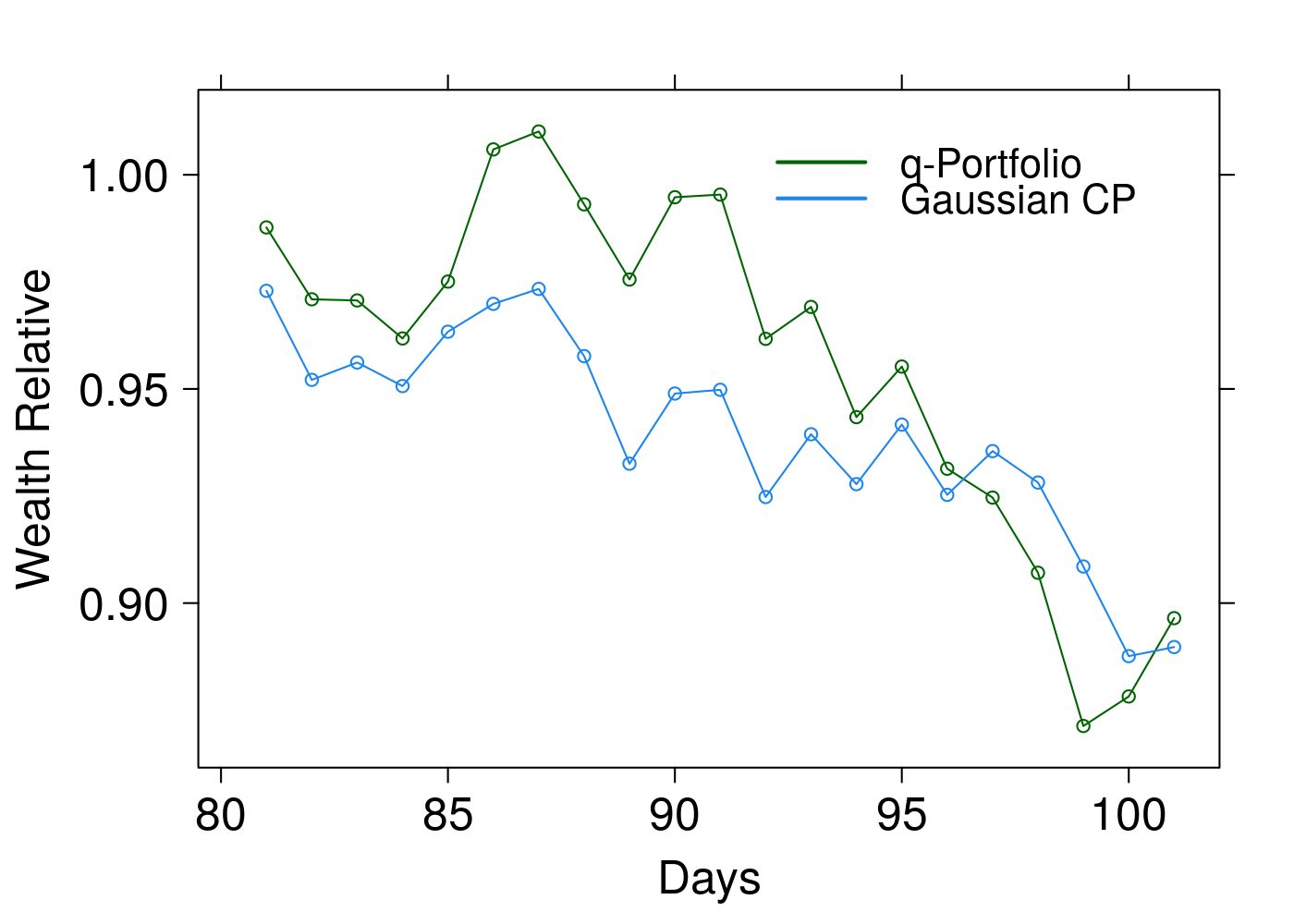} }

\vspace{-0.7cm}\hspace{1.6cm}(c) \hspace{8.7cm} (d)

\hspace{2.3cm} \resizebox{0.7\textwidth}{!}{ \includegraphics{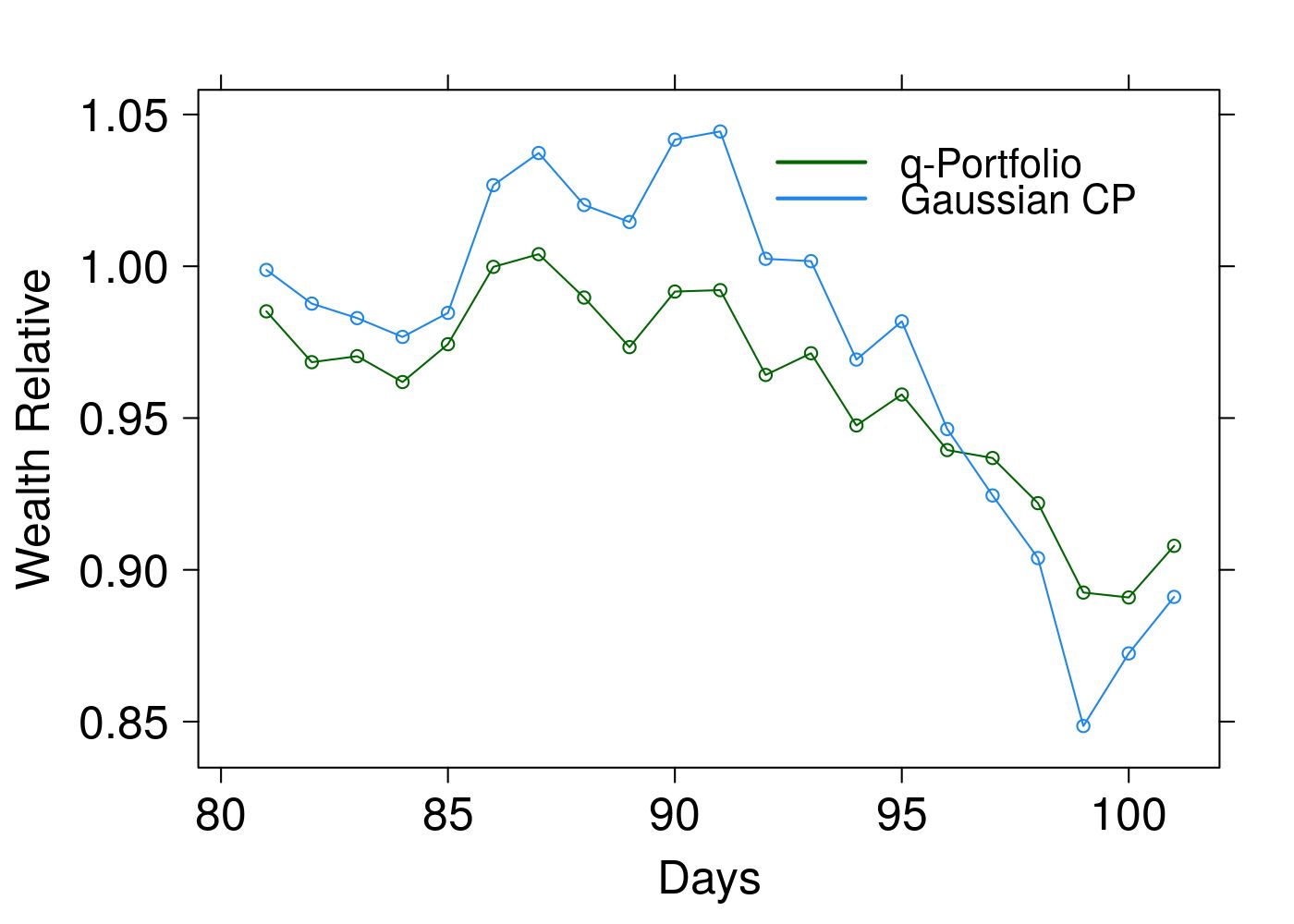} }

\vspace{-0.7cm}\hspace{6.4cm} (e) 
 
\caption{Wealth relative for the $q$-portfolio and the Gaussian Cover's Portfolio (CP) in the period 05/01/2018 to 05/31/2018, for three stocks: a) PETR4, VALE3 and BBDC4, b) PETR4, VALE3 and BBAS3, c) BBAS3, BBDC4 and VALE3, d) BBAS3, BBDC4 and PETR4.; and four stocks: e) BBAS3, BBDC4, PETR4 and VALE3.   } \label{fig-3}
\vspace{-1.0cm}
\end{figure}

After maximization, we apply the acquired portfolio for the next month 05/01/2018 to 05/31/2018, that represent 21 days of movement in $[B]^{3}$,  and calculate the wealth relative for two stocks as present in figure \ref{fig-2}, and for three and four stocks as shows in figure \ref{fig-3}. Analyzing this figures we can observe that the q-portfolio brings a better results that the Gaussian Cover's portfolio. 
Indeed the q-portfolio presents a higher wealth relative in $67,10 \%$ of all days, achieving $75,00 \%$ for three stocks set.

We present the results of wealth relative to two, three and four stocks for the period in table \ref{tab-2}, which reinforce that wealth relative for $q$-portfolio presents higher value at the end of the period in almost all situations, except for BBAS3 and PETR4 case.

\begin{table}[t]
\small
\setlength\tabcolsep{2pt} 
\caption{\label{tab-2}Wealth relative for Gaussian Cover's Portfolio (CP) and $q$-portfolio theory for two, three and four stocks in the period between 05/01/2018 to 05/31/2018.}
\hspace{-2.1cm}\begin{tabular}{cccccccccccc}
\hline\noalign{\smallskip} 

\multirow{4}{*}{Stocks} & \multirow{2}{*}{BBAS3}& \multirow{2}{*}{BBAS3} & \multirow{2}{*}{BBAS3} & \multirow{2}{*}{BBDC4} & \multirow{2}{*}{BBDC4} &  \multirow{2}{*}{PETR4}  & \multirow{1}{*}{BBAS3} & \multirow{1}{*}{BBDC4} & \multirow{1}{*}{BBAS3} & \multirow{1}{*}{BBAS3} & BBAS3  \\ 

 &   &   &   &  &  &  & \multirow{2}{*}{PETR4} & \multirow{2}{*}{PETR4} & \multirow{2}{*}{BBDC4} & \multirow{2}{*}{BBDC4} & BBDC4 \\ 

& \multirow{2}{*}{BBDC4} & \multirow{2}{*}{PETR4} & \multirow{2}{*}{VALE3} & \multirow{2}{*}{PETR4} & \multirow{2}{*}{VALE3} &  \multirow{2}{*}{VALE3}& \multirow{3}{*}{VALE3} & \multirow{3}{*}{VALE3} & \multirow{3}{*}{PETR4} & \multirow{3}{*}{VALE3} & PETR4  \\ 

 &   &   &   &  &   &  &   &  &   &  & VALE3 \\  \hline\noalign{\smallskip} 

\hspace{-0.4cm} Wealth relative \hspace{-0.4cm} & \multirow{2}{*}{$0.9010 $} &  \multirow{2}{*}{$0.9220 $} &  \multirow{2}{*}{$0.9354 $} &  \multirow{2}{*}{$0.8870 $} &  \multirow{2}{*}{$0.9008 $} &  \multirow{2}{*}{$0.8762 $} &  \multirow{2}{*}{$0.8765 $} &  \multirow{2}{*}{$0.8996 $} &  \multirow{2}{*}{$0.9008 $} &  \multirow{2}{*}{$0.9008 $} &  \multirow{2}{*}{$0.8896 $}    \\

Gaussian CP           &  &  &  &  &   &  &  &  &  &  &  \\  \hline\noalign{\smallskip} 

\hspace{-0.4cm} Wealth relative \hspace{-0.4cm}  & \multirow{2}{*}{$0.9134 $} &  \multirow{2}{*}{$0.9021 $} &  \multirow{2}{*}{$0.9495 $} &  \multirow{2}{*}{$0.8905 $} &  \multirow{2}{*}{$0.9394 $} &  \multirow{2}{*}{$0.9278 $} &  \multirow{2}{*}{$0.9196 $} &  \multirow{2}{*}{$0.9120 $} &  \multirow{2}{*}{$0.9012 $} &  \multirow{2}{*}{$0.9346 $} &  \multirow{2}{*}{$0.9142 $} \\

$q$-Portfolio           &  &  &  &  &   &  &  &  &  &  &  \\  \hline\noalign{\smallskip} 

\hspace{-0.4cm} $q$-Wealth relative \hspace{-0.4cm}  & \multirow{2}{*}{$0.9140 $} &  \multirow{2}{*}{$0.9022 $} &  \multirow{2}{*}{$0.9490 $} & \multirow{2}{*}{$0.8917 $} &  \multirow{2}{*}{$0.9397 $} &  \multirow{2}{*}{$0.9269 $} &  \multirow{2}{*}{$0.9174 $} &  \multirow{2}{*}{$0.9137 $} & \multirow{2}{*}{$0.9018 $} &  \multirow{2}{*}{$0.9348 $} &  \multirow{2}{*}{$0.9146 $}   \\

$q$-Portfolio           &  &  &  &  &   &  &  &  &  &  &  \\  \hline\noalign{\smallskip} 

\multirow{2}{*}{$q$-value}  & \multirow{2}{*}{$1.6278 $} &  \multirow{2}{*}{$ 1.6413$} &  \multirow{2}{*}{$1.6050 $} &  \multirow{2}{*}{$1.5888 $} &  \multirow{2}{*}{$1.5960 $} &  \multirow{2}{*}{$1.5871 $} &  \multirow{2}{*}{$1.4885 $} &  \multirow{2}{*}{$1.4763 $} &  \multirow{2}{*}{$1.4944 $} &  \multirow{2}{*}{$1.4830 $} &  \multirow{2}{*}{$1.3951 $}   \\

 &  &  &  &  &   &  &  &  &  &  &  \\  \hline\noalign{\smallskip}

\end{tabular}
\end{table}

At end of the period of 21 days of movement, we can see that wealth relative for $q$-Portfolio brings better results that the Gaussian Cover's portfolio in ten of the eleven cases analyzed.
In eight of the eleven cases the q-portfolio begin with higher values that the Gaussian Cover's portfolio, brings better results at the end of the period, showing that our method deals better with Brazilian stock market $[B]^{3}$ data compared to the Gaussian Cover's portfolio theory, being  able to predict a higher wealth relative value.

The parameter $q$ in all cases is between $1.39$ to $1.65$ closer to value obtained in \cite{TsBor,Borland}, indicating that  behavior followed by the stock market is well represented by this range of parameters in Tsallis statistics.
In the same way we can apply our formalism to the case with several stocks and analyze the portfolio of many sizes.

 A measure of the portfolio's risk-adjusted return can be realized by calculating the Sharpe ratio
\begin{eqnarray}
 S_{a} & = & \displaystyle \frac{E \left[ R_{t} - R_{f}  \right]  }{ \sigma \left[ R_{t} \right] } 
\end{eqnarray}
 and the Sortino ratio
\begin{eqnarray}
  S_{o} & = & \displaystyle \frac{ E \left[ R_{t} - T \right] }{TDD}
\end{eqnarray}
where $R_{t}$ is portfolio's return, $R_{f}$ denotes the risk-free rate of return, $\sigma \left[ R_{t} \right]$ is the standard deviation of the portfolio return, $T$ denotes the target or require of return for an investment and $TDD = \sqrt{\frac{1}{N} \sum_{i=1}^{N}  min(0,R_{i} - T)^{2}  }$ is the target downside deviation.

We use a risk free rate and a  require rate of returns  $R_{f}= T = 0$ and calculate the Sharpe and Sortino ratio for the q-Portfolio and the Gaussian Cover's Portfolio. The results are present in table \ref{tab-s} and for $63,6\%$ of the cases the q-Portfolio performs better that the Gaussian Cover's Portfolio for both, Sharpe Ratio and Sortino Ratio.

\begin{table}[!ht]
\small 
\setlength\tabcolsep{2pt} 
\caption{\label{tab-s}Sharpe Ratio and Sortino ratio for Gaussian Cover's Portfolio (CP) and $q$-portfolio theory for two, three and four stocks.}
\hspace{-2.1cm}\begin{tabular}{cccccccccccc}
\hline\noalign{\smallskip} 

\multirow{4}{*}{Stocks} & \multirow{2}{*}{BBAS3}& \multirow{2}{*}{BBAS3} & \multirow{2}{*}{BBAS3} & \multirow{2}{*}{BBDC4} & \multirow{2}{*}{BBDC4} &  \multirow{2}{*}{PETR4}  & \multirow{1}{*}{BBAS3} & \multirow{1}{*}{BBDC4} & \multirow{1}{*}{BBAS3} & \multirow{1}{*}{BBAS3} & BBAS3  \\ 

 &   &   &   &  &  &  & \multirow{2}{*}{PETR4} & \multirow{2}{*}{PETR4} & \multirow{2}{*}{BBDC4} & \multirow{2}{*}{BBDC4} & BBDC4 \\ 

& \multirow{2}{*}{BBDC4} & \multirow{2}{*}{PETR4} & \multirow{2}{*}{VALE3} & \multirow{2}{*}{PETR4} & \multirow{2}{*}{VALE3} &  \multirow{2}{*}{VALE3}& \multirow{3}{*}{VALE3} & \multirow{3}{*}{VALE3} & \multirow{3}{*}{PETR4} & \multirow{3}{*}{VALE3} & PETR4  \\ 

 &   &   &   &  &   &  &   &  &   &  & VALE3 \\  \hline\noalign{\smallskip}

 \hspace{-0.4cm} Sharpe Ratio \hspace{-0.4cm}  & 
\multirow{2}{*}{$0.1444  $} &  \multirow{2}{*}{$0.8016  $} &  \multirow{2}{*}{$0.9042  $} &  
\multirow{2}{*}{$1.2921  $} &  \multirow{2}{*}{$0.1398  $} &  \multirow{2}{*}{$1.7122  $} &  
\multirow{2}{*}{$1.7122  $} &  \multirow{2}{*}{$0.1398  $} &  
\multirow{2}{*}{$0.1398  $} &  \multirow{2}{*}{$0.1398  $} &  
\multirow{2}{*}{$1.5155  $} \\

Gaussian CP            &  &  &  &  &   &  &  &  &  &  &  \\  \hline\noalign{\smallskip}

 \hspace{-0.4cm} Shape Ratio \hspace{-0.4cm} & 
\multirow{2}{*}{$0.5707  $} &  \multirow{2}{*}{$1.3652  $} &  \multirow{2}{*}{$0.9919  $} & 
\multirow{2}{*}{$1.0694  $} &  \multirow{2}{*}{$0.5666  $} &  \multirow{2}{*}{$1.5584  $} &  
\multirow{2}{*}{$1.4712  $} &  \multirow{2}{*}{$1.2452  $} &  
\multirow{2}{*}{$1.0881  $} &  \multirow{2}{*}{$0.7198  $} & 
\multirow{2}{*}{$0.9869  $}    \\

q-Portfolio          &  &  &  &  &   &  &  &  &  &  &  \\  \hline\noalign{\smallskip}

\hspace{-0.4cm} Sortino Ratio \hspace{-0.4cm}  & 
\multirow{2}{*}{$0.1889 $} &  \multirow{2}{*}{$1.3411 $} &  \multirow{2}{*}{$1.5606 $} &  
\multirow{2}{*}{$2.0246 $} &  \multirow{2}{*}{$0.1826 $} &  \multirow{2}{*}{$2.9201 $} &  
\multirow{2}{*}{$2.9201 $} &  \multirow{2}{*}{$0.1827 $} & 
\multirow{2}{*}{$0.1827 $} &  \multirow{2}{*}{$0.1826 $} &  
\multirow{2}{*}{$2.5377 $}   \\

Gaussian CP          &  &  &  &  &   &  &  &  &  &  &  \\  \hline\noalign{\smallskip} 

\hspace{-0.4cm} Sortino Ratio \hspace{-0.4cm}   & 
\multirow{2}{*}{$0.8820 $} &  \multirow{2}{*}{$2.3338 $} &  \multirow{2}{*}{$1.7462 $} &  
\multirow{2}{*}{$1.6042 $} &  \multirow{2}{*}{$0.8478 $} &  \multirow{2}{*}{$2.7156 $} &  
\multirow{2}{*}{$2.5691 $} &  \multirow{2}{*}{$2.0111 $} &  
\multirow{2}{*}{$1.737 $} &  \multirow{2}{*}{$1.1545 $} &  
\multirow{2}{*}{$1.5813 $}   \\

$q$-Portfolio    &  &  &  &  &   &  &  &  &  &  &  \\  \hline\noalign{\smallskip}

\end{tabular}
\end{table}

The values from table \ref{tab-s} indicating that q-Portfolio present a better risk-adjusted performance considering the total volatility and the downside risk, being able to generate a higher return that the Gaussian Cover's Portfolio.

\section{Conclusions \label{sec4} }
The Tsallis statistic allows us to generalize usual concepts through a non-extensive parameter $q$. At the limit $q \rightarrow 1$, the usual expressions are recovered.  Examples of interest include deformations of functions and algebraic operations. We explore these tools in Cover's approach from information theory to portfolio theory. In this way we define the growth $q$-rate $W_{q}$ of a stock market portfolio, the optimal growth $q$-rate $W_{q}^{*}$, $q$-wealth after n days using the portfolio $b^{*}$ and we show that the $q$-wealth after $n$ days with the optimal portfolio is given by the $q$-exponential function. In the context of causal portfolio we studied the asymptotic optimality and we derive that for, almost every sequence from stock market the optimal $q$-wealth after n days is greater than the $q$-wealth of any investor. It is important to note that the parameter $q$ establishes a correlation between stock prices on each day. In fact, by equation (\ref{qGau}), we can note that the proposed $q$-Gaussian is not a product of $q$-Gaussians, but it is a $q$-product. Based on this approach a Brazilian stock market analysis was performed in a small number of stocks. We show that $q$-wealth is better suited to empirical data than standard wealth and the optimal value for the non-extensive parameter $q$  between $1.39$ to $1.65$.  An analysis of the Sharpe ratio and Sortino ratio also ratifies advantages in using q-Portfolio.
As perspectives we intend to investigate i.i.d. markets to time-dependent market processes in this scenario. Furthermore, we pretend apply this formulation to analyze the Post-Modern Portfolio Theory exploring the nonextensive version of Downside Risk and of log-normal distribution.  

%




\begin{thebibliography}{00}


\bibitem{Mark}{\href{https://www.math.ust.hk/\~maykwok/courses/ma362/07F/markowitz_JF.pdf}{H. Markowitz, Portfolio Selection The Journal of Finance 7 (1952) 77-91.  }  }

\bibitem{Sharpe}{W. Sharpe, G. Alexander and J. Bailey, Investments, Prentice-Hall, New Jersey, 1985.}

\bibitem{Cover}{T. Cover, J. Thomas, Elements of Information theory, Wiley-Interscience, New York, 2006.}


\bibitem{Cover1}{\href{https://doi.org/10.1142/9789814293501_0012}{R. Bell, T. Cover, Competitive optimality of logarithmic investment, Math. Oper. Res.  5 (1980) 147-152.}}


\bibitem{Cover2}{\href{https://doi.org/10.1287/mnsc.34.6.724}{R. Bell, T. Cover, Game-theoretic Optimal Portfolios, Manage. Sci. 34 (1988) 724-733.}}


\bibitem{Cover3}{\href{https://doi.org/10.1142/9789814293501_0013}{A. Barron, T. Cover, A bound on the financial value of information, IEEE Trans. Inf. Theory 34 (1988) 1097-1100.}}


\bibitem{Kelly}{\href{https://doi.org/10.1002/j.1538-7305.1956.tb03809.x}{J. Kelly, A new interpretation of information rate, Bell Syst. Tech. J 35 (1956) 917-926.}}


\bibitem{Cover4}{\href{https://projecteuclid.org/euclid.aop/1176991793}{P. Algoet, T. Cover, Asymptotic optimality and asymptotic equipartition property of log-optimal investment, Ann. Prob. 16 (1988) 876-898.}}


\bibitem{Cover5}{\href{https://doi.org/10.1142/9789814293501_0015}{T. Cover, Universal portfolios, Math. Finance  1 (1991) 181-209.}}

\bibitem{theil65}{\href{http://www.jstor.org/stable/2351063}{ H. Theil, C. Leenders, Tomorrow on the Amsterdam Stock Exchange, Journal of Business 38 (1965) 277-284. }}

\bibitem{fama65b}{\href{http://www.jstor.org/stable/2351064}{ E. F. Fama, Tomorrow on the New York Stock Exchange, Journal of Business 38 (1965) 285-299. }}



\bibitem{ahmadi11}{\href{https://ieeexplore.ieee.org/document/6033932}{ A. Ahmadi-Javid, An information theoretic approach to constructing coherent risk measures, Proceedings of IEEE International Symposium on Information Theory, St. Petersburg, (2011)  2125-2127. }}

\bibitem{ahmadi12}{\href{https://doi.org/10.1007/s10957-011-9968-2}{ A. Ahmadi-Javid, Entropic value-at-risk. A new coherent risk measure, J. Optim. Theory Appl. 155 (2012) 1105-1123. }}


\bibitem{ahmadi19}{\href{https://doi.org/10.1016/j.ejor.2019.02.007}{ A. Ahmadi-Javid, M. Fallah-Tafti, Portfolio optimization with entropic value-at-risk, European Journal of Operational Research 279 (2019) 255-241. }}

\bibitem{memmel03}{\href{https://ssrn.com/abstract=412588}{  C. Memmel, Performance hypothesis testing with the Sharpe ratio, Finance Letters 1 (2003) 21-23.}}

\bibitem{ling19}{\href{https://doi.org/10.1016/j.ejor.2019.01.012}{  A. Ling, J. Sun, M. Wang, Robust multi-period portfolio selection based on downside risk with asymmetrically distributed uncertainty set, European Journal of Operational Research (2019) 1-36.}}


\bibitem{St1}{R. Mantegna, H. Stanley, An Introduction to Econophysics: Correlations and Complexity in Finance, Cambridge University Press, Cambridge, 2000.}


\bibitem{Bou}{J. Bouchaud, M. Potters, Theory of Financial Risk and Derivative Pricing, Cambridge University Press, Cambridge, 2003.}

\bibitem{Isard}{\href{https://www.jstor.org/stable/1884452}{W. Isard, Location theory of trade theory: short-run analysis, Quarterly Journal of Economics 2 (1954) 305-320.}}

\bibitem{Zu1}{\href{https://doi.org/10.1016/j.physa.2010.01.007}{L. Zunino, M. Zanin, B.M. Tabak, D.G. Pérz, O. Rosso, Complexity-entropy causality plane: A useful approach to quantify the stock market inefficiency, Physica A 389 (2010) 1891-1901.}}


\bibitem{Zu2}{\href{https://doi.org/10.1016/j.physa.2012.04.009} {L. Zunino, A.F. Bariviera, M.B. Guercio, L.B. Martinez, O.A. Rosso, On the efficiency of sovereign bond markets, Physica A 391 (2012) 4342-4349}}

\bibitem{zunino16}{\href{http://doi.org/10.1016/j.physa.2016.03.007}{  L. Zunino, A. Bariviera, M. Guercio, L. Martinez, O. Rosso, Monitoring the informational efficiency of European corporate bond markets with dynamical permutation min-entropy, Physica A 456 (2016) 1-9.}}



\bibitem{Tsallis2}{C. Tsallis, Introduction to nonextensive statistical mechanics: Approaching a Complex World, Springer-verlag, New York, 2009.}


\bibitem{Tsallis3}{M. Gell-Mann, C. Tsallis, Nonextensive Entropy: Interdisciplinary Applications, Oxford University Press, Oxford, 2003.}


\bibitem{Tsallis1}{\href{https://doi.org/10.1007/BF01016429}{C. Tsallis, Possible generalization of Boltzmann-Gibbs statistics, J. Stat. Phys. 52 (1988) 479-487.}}

\bibitem{Tsallis4}{\href{https://doi.org/10.1088/1742-5468/2008/09/P09006}{A. Rodr\'iguez, V. Schw\"ammle, C. Tsallis, Strictly and asymptotically scale invariant probabilistic models of N correlated binary random variables having $q$-Gaussians as $N\rightarrow \infty$ limiting distributions,  J. Stat. Mech. (2008) P09006. }}



\bibitem{TsBor}{\href{https://doi.org/10.1016/S0378-4371(03)00042-6}{ C. Tsallis, C. Anteneodo, L. Borland, R. Osorio, Nonextensive statistical mechanics and economics, Physica A 324 (2003) 89-100.}}


\bibitem{Borland}{\href{https://doi.org/10.1103/PhysRevLett.89.098701}{L. Borland, Option Pricing Formulas Based on a Non-Gaussian Stock Price Model, Physical Review Letters 89 (2002)  098701.}}


\bibitem{mandelbrot63}{\href{https://doi.org/10.1007/978-1-4757-2763-0_14}{B. Mandelbrot, The variation of certain speculative Prices, J. Business, 39 (1963) 394-419.}}

\bibitem{fama65}{\href{https://www.jstor.org/stable/2350752?casa_token=l0i1pyuHfikAAAAA:cyDgakLiBrngXvP4Aq4Y8uJxy2ApgIltQ3KFExWsu0SnMpfCw4R2PoLhcpBH6pl6eUjadw5HsEIjb6y2FxXQQTgGAdH4fGwef3tnMDw71G39mNZRXTvc&seq=1#metadata_info_tab_contents}{E. Fama, The Behavior of stock-market prices, J. Business 38 (1965) 34-105.}}



\bibitem{Borges}{\href{https://doi.org/10.1016/j.physa.2004.03.082}{E. Borges, A possible deformed algebra and calculus inspired in nonextensive thermostatistics, Physica A 340 (2004)  95-101.}}

\bibitem{Umarov1}{\href{https://doi.org/10.1063/1.2828756}{S. Umarov, C. Tsallis, Multivariate Generalizations of the q-Central Limit Theorem Consistent With Nonextensive Statistical Mechanics, AIP Conference Proeedings 965 (2007) 34-42.}}



\bibitem{borland2004a}{\href{https://www.tandfonline.com/doi/abs/10.1080/14697680400008684}{L. Borland, J.P. Bouchaud, A non-Gaussian option pricing model with skew, Quantit. Finance 4 (2004) 499-514.}}


\bibitem{queiros2005a}{\href{https://iopscience.iop.org/article/10.1209/epl/i2004-10436-6/meta}{S.M.D. Queiros, C. Tsallis, Bridging a paradigmatic financial model and nonextensive entropy, Europhys. Lett. 69 (2005) 893-899.}}

\bibitem{vellekoop2007}{\href{https://doi.org/10.1080/14697680601077967}{M. Vellekoop, H. Nieuwenhuis, On option pricing models in the presence of heavy tails, Quantit. Finance 7 (2007) 563-573.}}


\bibitem{drozdz2007b}{\href{https://doi.org/10.1016/j.physa.2007.04.130}{S. Dro\.{z}d\.{z} et. al., Stock market return distributions: From past to present, Physica A 383 (2007) 59-64.}}


\bibitem{pan2008}{\href{https://doi.org/10.1016/j.physa.2007.11.031}{R. Pan, S. Sinha, Inverse-cubic law of index fluctuation distribution in Indian markets, Physica A 387 (2008) 2055-2065.}}


\bibitem{eryigit2009}{\href{https://doi.org/10.1016/j.physa.2009.01.019}{M. Eryi\u{g}it, S. \c{C}ukur, R. Eryi\u{g}it, Tail distribution of index fluctuations in World markets, Physica A 388 (2009) 1879-1886.}}


\bibitem{gopikrishnana1998}{\href{https://doi.org/10.1007/s100510050292}{P. Gopikrishnan, Inverse cubic law for the distribution of stock price variations, Eur. Phys. J. B, 3 (1998) 139-140.}}


\bibitem{drozdz2007a}{\href{https://doi.org/10.1016/j.physa.2006.07.035}{R. Rak, S. Dro\.{z}d\.{z}, J. Kwapie\'{n}, Nonextensive statistical features of the Polish stock market fluctuations, Physica A {374 (2007) 315-324.}}


\bibitem{ruiz2018}{\href{https://doi.org/10.1140/epjb/e2017-80535-3}{G. Ruiz, A. de Marcos, Evidence for criticality in financial data, Eur. Phys. J. B, 91 (2018) 1-5.}}


\bibitem{Durrett}{R. Durrett, Probability: Theory and Examples, Cambridge Series in Statistical and Probabilistic Mathematics, Cambridge, 2010.}


\bibitem{r-cran}{\href{http://www.R-project.org/}{R Core Team, R: A language and environment for statistical computing, R Foundation for Statistical Computing, Viena, 2013.}}


\bibitem{gethfdata}{\href{http://dx.doi.org/10.2139/ssrn.2824058}{M. Perlin, H. Ramos, GetHFData: A R Package for Downloading and Aggregating High Frequency Trading Data from Bovespa, Brazilian Review of Finance  14 (2016) 1-33.}}


\bibitem{deoptim}{\href{https://journal.r-project.org/archive/2011/RJ-2011-005/index.html}{D. Ardia, K. Boudt, P. Carl, K. Mullen, B. Peterson, Differential Evolution with DEoptim, The R Journal  3 (2011) 27-34.}}


\bibitem{diffevol}{K. Price, R. Storn, J. Lampinen, Differential Evolution - A Practical Approach to Global Optimization, Springer-Verlag, Berlin, 2006.}}


\bibitem{pracma}{\href{https://cran.r-project.org/web/packages/pracma/index.html}{H. Borchers, pracma: Practical Numerical Math Functions, R package  2.1.8 (2018).}}


\bibitem{becs}{W. H\"ardle, O. Okhrin, Y. Okhrin, Basic Elements of Computational Statistic, Springer International Publishing, Berlin, 2017.}


\bibitem{cubature}{\href{https://cran.r-project.org/web/packages/cubature/index.html}{B. Narasimhan, S. Johnson, cubature: Adaptive Multivariate Integration over Hypercubes, R package  1.4 (2018). }}




\bibitem{muller88}{\href{https://doi.org/10.2143/AST.18.2.2014947}{H. Muller, Modern Portfolio Theory: Some Main Results, The Journal of the IAA, 18 (1988) 127-145. }}


\bibitem{inuiguchi2000}{\href{https://doi.org/10.1016/S0165-0114(99)00026-3}{M. Inuiguchi, T. Tanino, Portfolio selection under independent possibilistic information, Fuzzy Sets and Systems, 115 (2000) 83-92.}}


\bibitem{li2013}{\href{https://doi.org/10.1145/2435209.2435213}{B. Li, et al., Confidence Weighted Mean Reversion Strategy for Online Portfolio Selection, ACM Transactions on Knowledge Discovery from Data (TKDD),7  (2013) 1-38.}}

\bibitem{tunc2013}{\href{https://doi.org/10.1109/TSP.2013.2258339}{S. Tunc, M. Donmez, S. Kozat, Optimal Investment Under Transaction Costs: A Threshold Rebalanced Portfolio Approach, IEEE Transactions on Signal Processing, 61  (2013) 3129-3142}.}


\bibitem{matesanz08}{\href{https://doi.org/10.1142/S0129183108012789}{D. Matesanz, G. Ortega, A (Econophysics) note on volatility in exchange
rate time series. Entropy as a ranking criterion, International Journal of Modern Physics C, 19 (2008) 1095-1103.}}

\bibitem{markowitz12}{\href{https://doi.org/10.1016/j.ejor.2012.08.023}{H. Markowitz, Mean–variance approximations to expected utility, European Journal of Operational Research, 234 (2014) 346-355.}}

\bibitem{zhao2018}{\href{http://doi.org/10.12693/APhysPolA.133.1170}{P. Zhao, J. Wang, Y. Song, Optimal Portfolio under Non-Extensive Statistical Mechanics and Value-at-Risk Constraints, Acta Physica Polonica A, 133  (2018) 1170-1173.}}


\end{thebibliography}


\end{document}